\documentclass{PoS}

\usepackage{graphicx}

\usepackage[english]{babel}
\usepackage[utf8]{inputenc}
\usepackage[T1]{fontenc}

\usepackage{amsfonts}
\usepackage{amssymb}

\usepackage{tikz}
\usepackage{nicefrac}

\usetikzlibrary{decorations.pathmorphing} 
\usetikzlibrary{decorations.pathreplacing}
\usetikzlibrary{decorations.markings}
\usetikzlibrary{arrows} 
\usepackage{multirow}

\definecolor{DRed}{RGB}{139,0,0}
\definecolor{DBlue}{RGB}{0,0,139}

\newcommand{\ee}{\,\mathrm{e}}

\title{Insights into the heavy dense QCD phase diagram using Complex Langevin simulations}

\ShortTitle{Insights into the heavy dense QCD phase diagram using Complex Langevin simulations}

\author{Gert Aarts$^{1}$, Felipe
Attanasio$^{1,2}$, \speaker{Benjamin Jäger}$^{1}$, Erhard Seiler$^{3}$, Dénes
Sexty$^{4}$, Ion-Olimpiu Stamatescu$^{5}$.\\
$^{1}$Department of Physics, College of Science, Swansea University, Swansea, 
UK\\
$^{2}$CAPES Foundation, Ministry of Education of Brazil, Brasília, Brazil\\
$^{3}$Max-Planck-Institut für Physik (Werner-Heisenberg-Institut), München,
Germany\\
$^{4}$Department of Physics, Bergische Universität Wuppertal, Wuppertal,
Germany\\
$^{5}$Institut für
Theoretische Physik, Universität Heidelberg, Heidelberg, Germany\\
E-mail: 
\email{B.Jaeger@swansea.ac.uk} 
}

\abstract{
Complex Langevin simulations provide an alternative to sample path integrals
with complex weights and therefore are suited to determine the phase diagram
of QCD from first principles. Adaptive step-size scaling and gauge cooling 
are used to improve the convergence of our simulations. We present results 
for the phase diagram of QCD in the limit of heavy quarks and discuss 
the order of the phase transitions, which are studied by varying 
the spatial simulation volume.
  }

\FullConference{The 33rd International Symposium on Lattice Field Theory\\
                 14 - 18 July  2015\\
                 Kobe International Conference Center, Kobe, Japan}

\begin{document}
  
\section{Introduction}
\begin{figure}[!ht]
\centering
\begin{tikzpicture}[scale=1.4]
\begin{scope}[>=latex]
\draw[black,->, ultra thick] (0.0,0.0) -- (0.0,3.20); 
\draw[black,->, ultra thick] (0.0,0.0) -- (5.0,0.00); 
\draw[black,very thick, dashed]  (3.0,0.0) .. controls (3,1.80) and (3,1.80) ..
(0,2.2);  
\draw[black,very thick, dashed]  (2.4,0.0) .. controls (2.3,0.60) and (2.3,0.60)
..
(2.2,0.65); 
\draw[black,very thick, dashed]  (3.4,0.0) .. controls (3.6,0.60) and
(3.6,0.60) ..(4.4,0.8); 
\node[below=0.1cm] at (2.5,0) {$\mu$};
\node[left=0.1cm] at (0,2) {$T$};
\node[left=0.1cm] at (2.0,1.0) {Hadrons};
\node[left=0.1cm] at (4.5,2.7) {Quark-Gluon};
\node[left=0.1cm] at (4.2,2.4) {Plasma};
\node[left=0.1cm] at (4.9,1.4) {Nuclear matter};
\node[left=0.1cm] at (5.0,0.55) {Colour};
\node[left=0.1cm] at (5.75,0.25) {Superconductor?};
\draw[black,->, thick] (3.2,1.2) -- (2.7,0.5); 
\filldraw [black] (1.7,1.95) circle (2.5pt);
\node[above=0.1cm] at (1.7,1.95) {Critical point?};
\end{scope} 
\end{tikzpicture}
\caption{A scenario of the QCD phase diagram.}
\label{Fig1}
\end{figure}
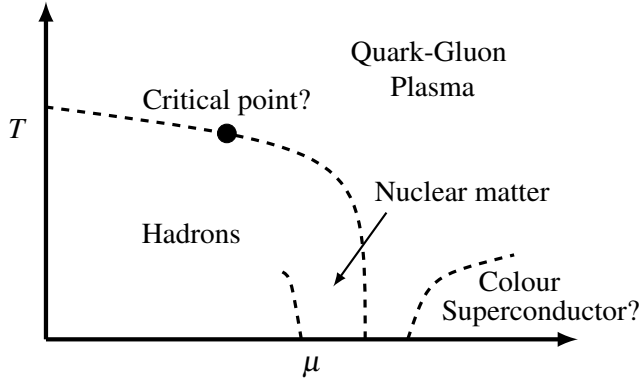

Complex Langevin simulations~[1-4]
have recently become an active field of research
with the development of \emph{gauge
cooling}~[5,\,6] 
and the prospect of enabling
studies of QCD with non-vanishing chemical potential~[7].
Here we report on our ongoing project studying the phase diagram of QCD. A
sketch of the expected phase diagram for QCD is shown in Figure\,\ref{Fig1}.
First principle calculations are a crucial input for understanding the
behaviour of strong interactions in heavy ion collisions and in neutron stars.

Before studying the QCD phase diagram of full QCD, we consider 
the heavy dense approximation of QCD
(HDQCD)~[4,\,5,\,8] 
as a first test and identify  the necessary step to determine the location and
order of the phase transitions. Currently, we do not consider higher orders in
the hopping parameter expansion, as done
in~[9-11], 
since our final goal is to study QCD with fully dynamical light quarks.

\section{Complex Langevin simulation}

Complex Langevin simulations are based on a stochastic process,
which is achieved by an update using simple matrix multiplication  
\begin{equation}
U_{\mu x}(t+\epsilon) = R(t)\, U_{\mu x}(t).
\end{equation}
The update matrix is composed of a deterministic drift and a stochastic term,
and can be written as
\begin{equation}
R(t) = \mathrm{exp}\left[ i \lambda \left(- \epsilon \, D_{U} S +
\sqrt{\epsilon}\, \eta \right) \right],\label{eqUpdate}
\end{equation}
where $\lambda$ are the Gell-Mann matrices of $\mathrm{SU(}3\mathrm{)}$ and
$D_{U} S$ is the gauge derivative of the action. The stochastic part is
provided by Gaussian white noise $\eta$. We use
Wilson's plaquette action in the gauge sector and include fermions by the
logarithm of the fermion determinant, which in the limit of heavy quarks can be
written for one flavour as
\begin{equation}
\det D(\mu) = \prod_{\vec{x}} \det \left( 1 + h\,\ee^{\mu/T} \,
\mathcal{P}_{\vec{x}} \right)^{2}\det \left( 1 + h\,\ee^{-\mu/T}\,
\mathcal{P}^{-1}_{\vec{x}} \right)^{2}
\end{equation} 
with $h = \left( 2 \, \kappa \right)^{N_\tau}$. Recent work~[12-15] 
has shown that the complex logarithm does not necessarily spoil Complex Langevin dynamics,
as long as the quark masses are sufficiently large. Due to the complex nature
of the QCD path integral in the presence of finite quark density, the gauge
links are an element of the gauge group $\mathrm{SL(}3,\mathbb{C}\mathrm{)}$.
We use adaptive step-size scaling~[16] 
and adaptive gauge
cooling~[5,\,6] 
to avoid large excursions into
the non-compact extension of $\mathrm{SU(}3\mathrm{)}$, which would invalidate
the justification of the approach~[17,\,18].
To check our simulations,
we monitor the 'unitnorm' during the Langevin evolution,
\begin{equation} 
d = \mathrm{Tr} \left( U U^{\dagger} - \mathbb{I} \right)^2, 
\end{equation}
which is a measure of the distance to the $\mathrm{SU(}3\mathrm{)}$ manifold
and a good indicator of stability of our runs.

\section{Numerical setup}

We study the phase diagram of heavy dense QCD using two flavours of heavy
quarks and work at a fixed lattice spacing (fixed $\beta$). To
improve our previous results~[19,\,20] 
we have
extended our simulations to cover two additional spatial volumes, i.e. $6^3$ and
$10^3$. We perform a simultaneous scan in the chemical potential and the temperature by varying $\mu$
and $N_\tau$. The range of our simulation parameter can be found in
Table\,\ref{tab}.
\begin{table}[h!]
\begin{center}
\begin{tabular}{|c|c|c|}
    	\hline
    	$\beta = 5.8$   & $N_f=2$ & $V=6^3,\, 8^3,\, 10^3$ \\
    	$\kappa = 0.04$ & $\mu = 0.0 - 3.2$ & $N_\tau = 2-32$\\
    	 & $a \sim 0.15\,\mathrm{fm}$ & $T = 670 - 42\,\mathrm{MeV}$\\
    	   	\hline
\end{tabular}
\caption{Summary of simulation parameters, where the chemical potential is
given in lattice units. The lattice spacing has been determined using the
Wilson flow in~[7,\,21]. 
}
\label{tab}
\end{center}
\end{table}
Our simulations have been extended to a maximum Langevin time of $500$, of 
which we discarded the first $100$ to remove thermalisation effects.
We use a step-size of $\epsilon \sim 10^{-4}$ and apply
adaptive step-size scaling to compensate for large forces in the Langevin
drifts. We determine observables every $10^{-2}$ Langevin time. Configurations
are typically  decorrelated when separated by approximately $10-100$ measurements,
depending on the actual value of the spatial volume, chemical potential and 
temperature. For each setup we have at least $4000 - 400$ decorrelated 
configurations. The auto-correlation has been determined using the
algorithm described in~[22].

\section{Results}

\begin{figure}[!ht]
	\centering
	\vspace{-0.5cm} 
	\begin{minipage}{0.95\linewidth}
	\centering
	\includegraphics[width=\linewidth]{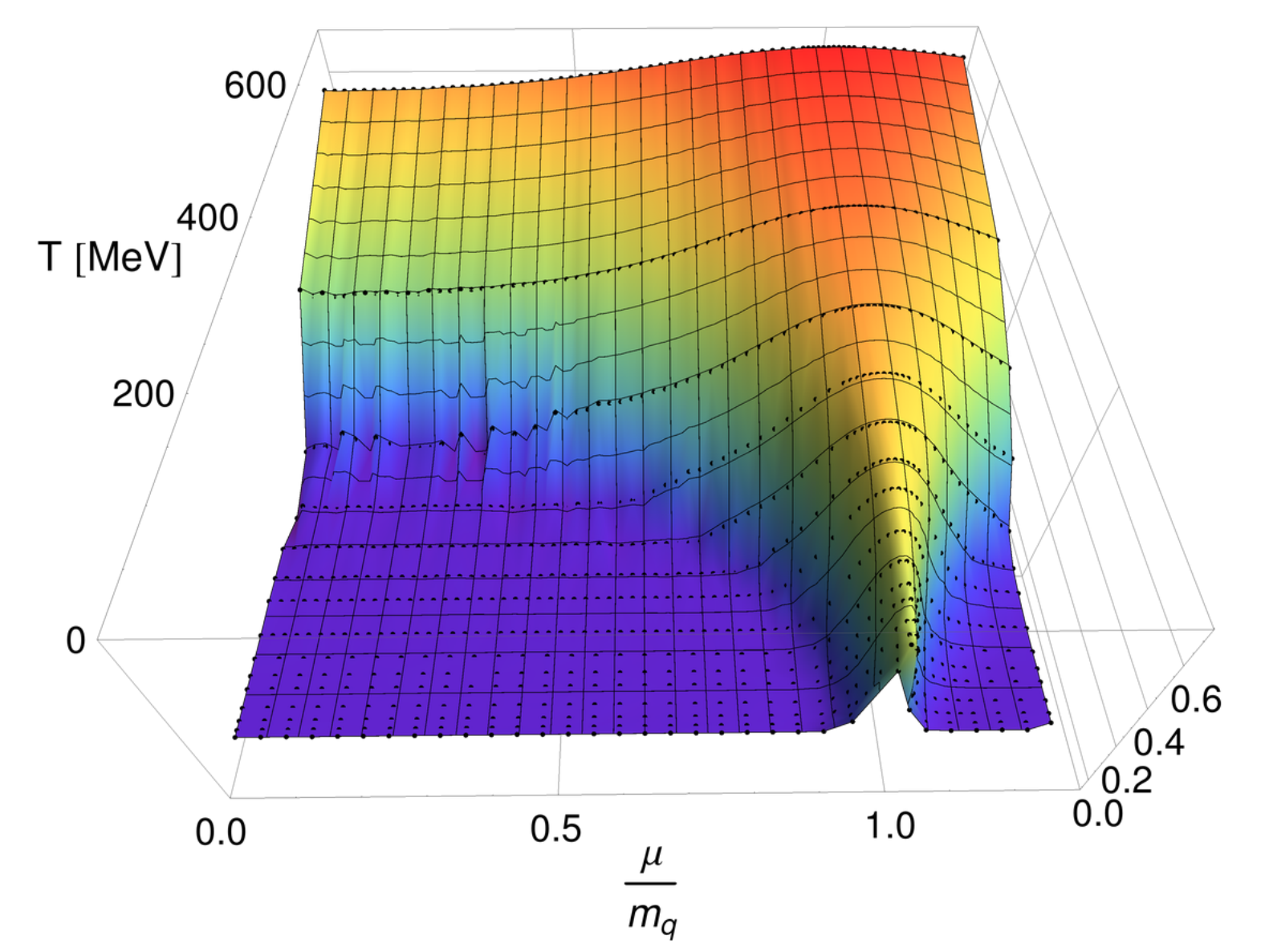}
	\end{minipage}
\caption{The Polyakov loop as function of temperature $T$ and chemical
potential $\mu$ for the $8^3$ ensembles. Each black dot represent the result of
a Complex Langevin simulation.}
\label{Fig2} 
\end{figure}
Figure\,2 shows the result for the Polyakov loop as a function of temperature
$T$ and the chemical potential $\mu$ for the spatial volume of $8^3$. The
temperature axis is shown in units of MeV using the lattice
spacing of $a\sim0.15\,\mathrm{fm}$, which has been obtained using the Wilson
flow~[7,\,21]. 
The chemical potential is shown in units of the quark mass, which for HDQCD can
be simply written as 
\begin{equation} 
m_q \equiv - \ln(2 \kappa) = 2.53 \sim \mu_c.
\end{equation}
The expected critical chemical potential $\mu_c$ is directly related to the
quark mass in the heavy dense approximation of QCD. The
Polyakov loop is an excellent quantity to study both transitions, i.e. the
deconfinement transition and the transition to higher densities. An intrinsic
lattice artefact is visible at large chemical potentials, which
can be understood by Pauli blocking. At high enough densities all lattice sites
are filled with fermions and no additional fermion can be added. The Polyakov
loop drops to zero in this unphysical regime, since the system is
effectively equivalent to a pure gauge. The fermion density
on the other hand shows saturation for high densities. Figure\,2 also shows a
cubic interpolation of the data, represented by the coloured surface connecting
the individual results to guide the eye. Figure\,3 shows the equivalent plot
for the Polyakov loop susceptibility as a function of $\mu$ and $T$ for 
our intermediate volume of $8^3$. This plot provides a good
representation of the phase boundary of HDQCD. The broadness seen for the
deconfinement transition is caused by the limited resolution in the temperature
direction, since the temporal extent is naturally an integer. Varying the gauge
coupling and thereby the lattice spacing will allow us to probe different
temperatures and study also the behaviour towards the continuum limit. 
\begin{figure}[!ht]   
	\centering  
	\hspace{-0.0cm} 
	\begin{minipage}{0.95\linewidth}
	\centering
	\includegraphics[width=\linewidth]{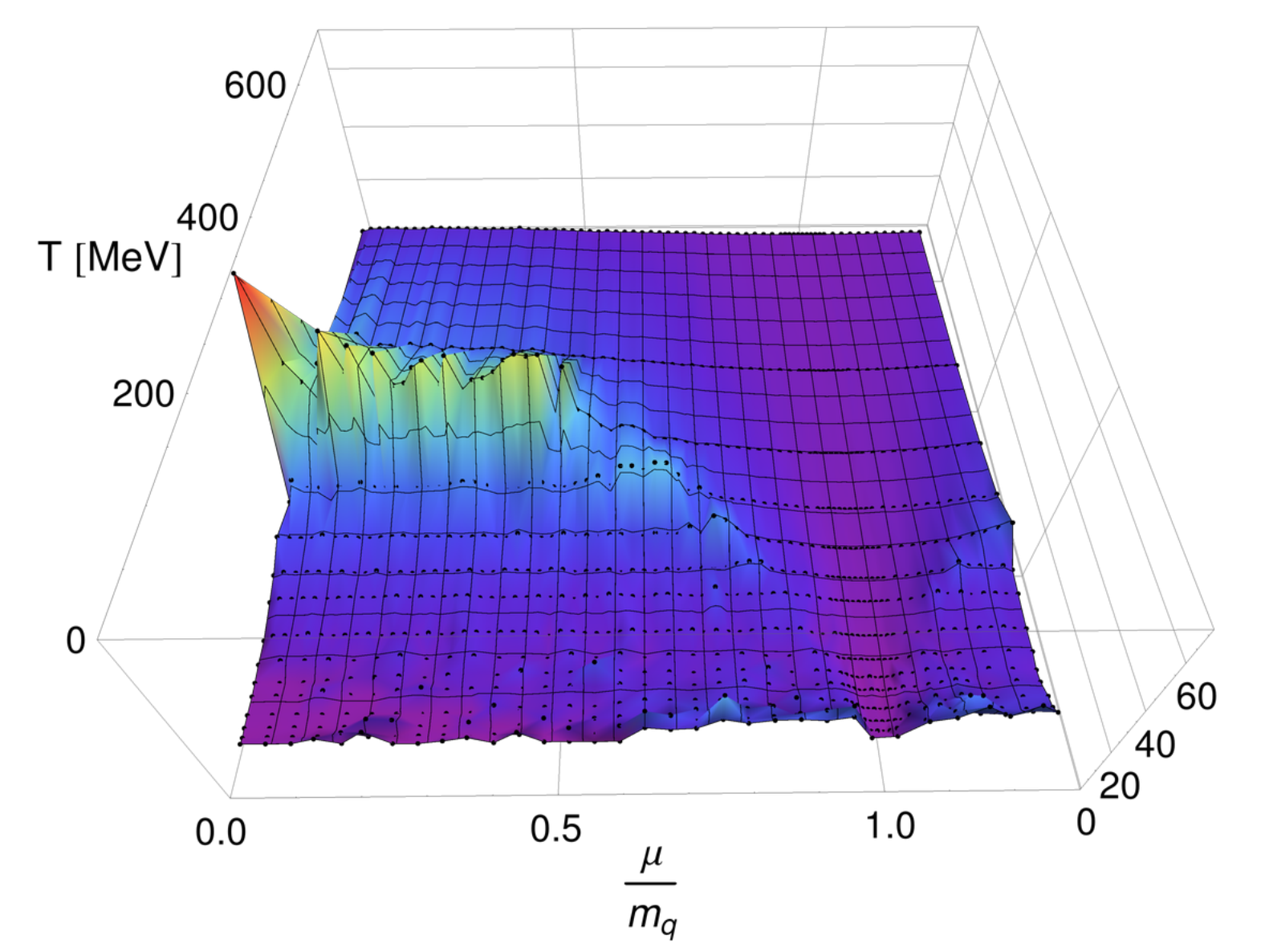}
	\end{minipage}   
\caption{The Polyakov loop susceptibility as function of $T$ and $\mu$ in the
same style as Figure\,2.}
\label{Fig3}
\end{figure}

To study the transitions in more detail, we have simulated HDQCD with three
different volumes, i.e. $6^3, 8^3$ and $10^3$.
Figure\,4 shows the susceptibility of the fermion density $n$, where the latter
is defined as
\begin{equation}
n = \frac{1}{N_\tau N_s^3} \frac{\partial\, \mathrm{ln}\, Z}{\partial \mu}.
\end{equation}
The fermion density and its susceptibility provide a very clear signal for the
transition to higher densities. An appoximate symmetry is visible in Figure\,4 
around the critical chemical potential $\mu_c$, which can be understood by
considering a symmetry between particles and holes~[23].
At half-filling, which is reached at $\mu_c$, the susceptibility drops and shows
a symmetric behaviour on both sides within statistical fluctuations. The upper
panel of Figure\,4 depicts the transition for one of our larger temperatures
of $T \sim 335\,\mathrm{MeV}$, whereas the lower panel shows the situation for a
smaller temperatures of $T \sim 167\,\mathrm{MeV}$.
\begin{figure}[!ht] 
	\centering 
	\hspace{-0.0cm} 
	\begin{minipage}{0.95\linewidth}
	\centering
	\includegraphics[width=\linewidth]{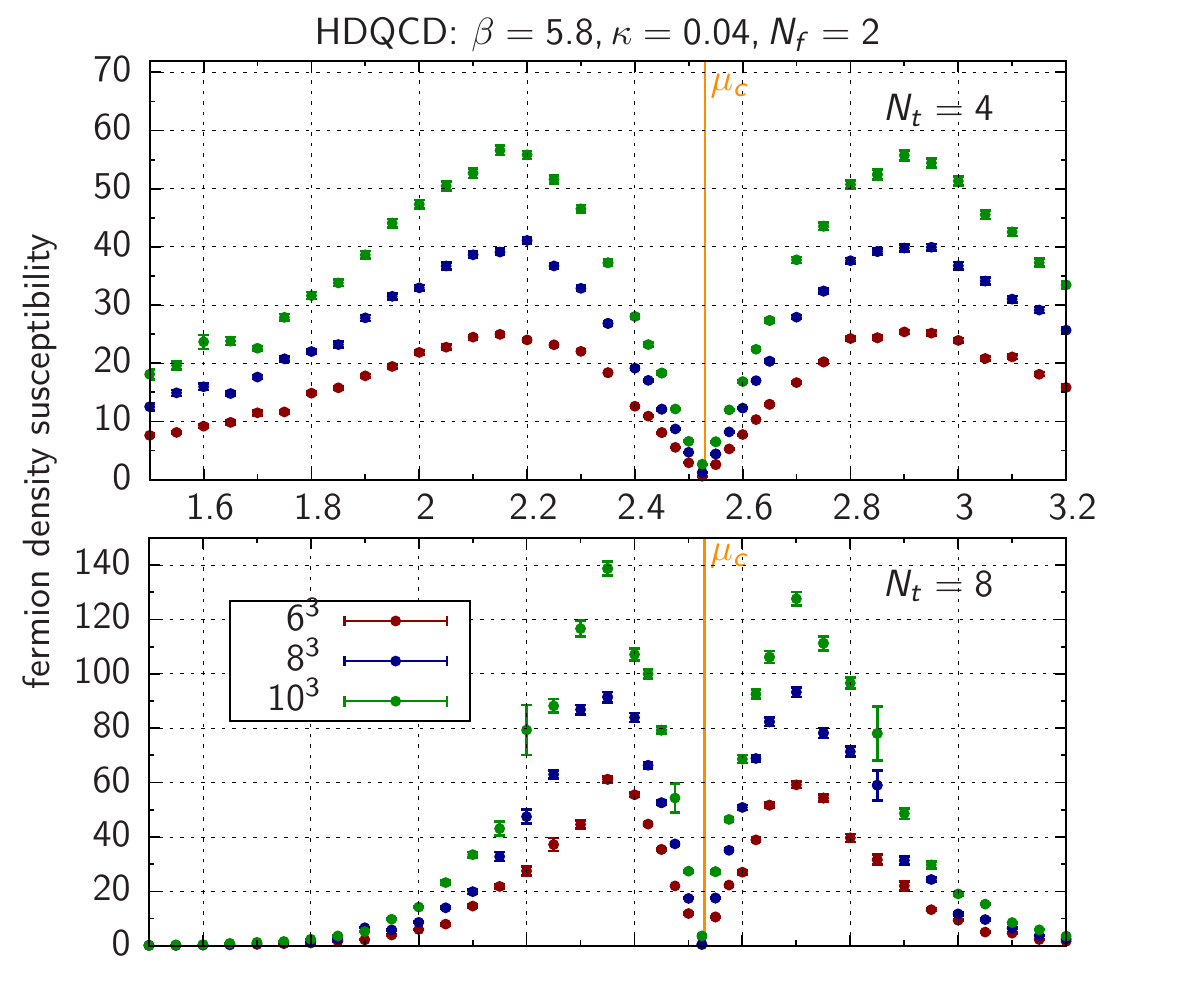}
	\end{minipage}   
\caption{The fermion density susceptibility as a function of the
chemical potential for two different temporal extents $N_t=4$ (upper) and
$N_t=8$ (lower). The volumes are shown in different colours. }
\label{Fig4}
\end{figure}

\section{Conclusions and Outlook}

Here we presented an update of our ongoing project to study the phase diagram
of (heavy dense) QCD from first principles using Complex Langevin simulations. We
find clear signals for the deconfinement transition and the transition to
higher density. Currently, we are extending our simulations to even larger
volumes around the transition, to better identify the order of the transitions.
Varying the gauge coupling will allow us to improve the resolution of
the deconfinement transition and study the continuum limit. With this study we
have presented the necessary steps and methods to study the phase diagram
of fully dynamical QCD using staggered
quarks~[7], 
which we plan to study next.
\newline \phantom{stuff}\newline

{\bf Acknowledgements:} 
We are grateful for the computing resources made available by HPC Wales. Part of this
work used the DiRAC BlueGene/Q Shared Petaflop system at the University of
Edinburgh, operated by the Edinburgh Parallel  Computing Centre on behalf of the
STFC DiRAC HPC  Facility (www.dirac.ac.uk). This equipment was  funded by BIS
National E-infrastructure capital grant ST/K000411/1, STFC capital grant
ST/H008845/1,  and STFC DiRAC Operations grants ST/K005804/1 and ST/K005790/1.
DiRAC  is part of the National E-Infrastructure.
We  acknowledge the STFC grant ST/L000369/1, the Royal Society and the Wolfson
Foundation. FA is grateful for the support through the Brazilian government programme “Science without Borders” under scholarship number BEX 9463/13-5.


\begin{thebibliography}{99}

\bibitem{Parisi:1984cs}
  G.~Parisi,
  Phys.\ Lett.\ B {\bf 131} (1983) 393.

\bibitem{Klauder:1983nn}
  J.~R.~Klauder,
  Acta Phys.\ Austriaca Suppl.\  {\bf 25} (1983) 251.

\bibitem{Klauder:1983sp}
  J.~R.~Klauder,
  Phys.\ Rev.\ A {\bf 29} (1984) 2036.

\bibitem{Aarts:2008rr}
  G.~Aarts and I.~O.~Stamatescu,
  JHEP {\bf 0809} (2008) 018.

\bibitem{Seiler:2012wz}
  E.~Seiler, D.~Sexty and I.~O.~Stamatescu,
  Phys.\ Lett.\ B {\bf 723} (2013) 213.

\bibitem{Aarts:2013uxa}
  G.~Aarts, L.~Bongiovanni, E.~Seiler, D.~Sexty and I.~O.~Stamatescu,
  Eur.\ Phys.\ J.\ A {\bf 49} (2013) 89.


\bibitem{Sexty:2013ica}
  D.~Sexty,
  Phys.\ Lett.\ B {\bf 729} (2014) 108.
 

\bibitem{Bender:1992gn}
  I.~Bender, T.~Hashimoto, F.~Karsch, V.~Linke, A.~Nakamura,  {\it et al.},
  Nucl.\ Phys.\ Proc.\ Suppl.\  {\bf 26} (1992) 323.

\bibitem{DePietri:2007ak}
  R.~De Pietri, A.~Feo, E.~Seiler and I.~O.~Stamatescu,
  Phys.\ Rev.\ D {\bf 76} (2007) 114501

\bibitem{Fromm:2011qi}
  M.~Fromm, J.~Langelage, S.~Lottini and O.~Philipsen,
  JHEP {\bf 1201} (2012) 042
  
\bibitem{Aarts:2014bwa}
  G.~Aarts, E.~Seiler, D.~Sexty and I.~O.~Stamatescu,
  Phys.\ Rev.\ D {\bf 90} (2014) 11,  114505
\bibitem{Mollgaard:2013qra}
  A.~Mollgaard and K.~Splittorff,
  Phys.\ Rev.\ D {\bf 88} (2013) 11,  116007
  
\bibitem{Splittorff:2014zca}
  K.~Splittorff,
  Phys.\ Rev.\ D {\bf 91} (2015) 3,  034507

\bibitem{Nishimura:2015pba}
  J.~Nishimura and S.~Shimasaki,
  Phys.\ Rev.\ D {\bf 92} (2015) 1,  011501

\bibitem{Greensite:2014cxa}
  J.~Greensite,
  Phys.\ Rev.\ D {\bf 90} (2014) 11,  114507

\bibitem{Aarts:2009uq}
  G.~Aarts, E.~Seiler and I.~O.~Stamatescu,
  Phys.\ Rev.\ D {\bf 81} (2010) 054508.

\bibitem{Aarts:2011ax}
  G.~Aarts, F.~A.~James, E.~Seiler and I.~O.~Stamatescu,
  Eur.\ Phys.\ J.\ C {\bf 71} (2011) 1756.
  
\bibitem{Aarts:2009dg}
  G.~Aarts, F.~A.~James, E.~Seiler and I.~O.~Stamatescu,
  Phys.\ Lett.\ B {\bf 687} (2010) 154.

\bibitem{Aarts:2014kja}
  G.~Aarts, F.~Attanasio, B.~Jäger, E.~Seiler, D.~Sexty and I.~O.~Stamatescu,
  PoS LATTICE {\bf 2014} (2014) 200
  
\bibitem{Aarts:2015yba}
  G.~Aarts, F.~Attanasio, B.~Jäger, E.~Seiler, D.~Sexty and I.~O.~Stamatescu,
  Acta Phys.\ Polon.\ Supp.\  {\bf 8} (2015) 2,  405


\bibitem{Borsanyi:2012zs}
  S.~Borsanyi, S.~Durr, Z.~Fodor, C.~Hoelbling, S.~D.~Katz,  {\it et al.},
  JHEP {\bf 1209} (2012) 010.
  
\bibitem{Wolff:2003sm}
  U.~Wolff [ALPHA Collaboration],
  Comput.\ Phys.\ Commun.\  {\bf 156} (2004) 143
   [Comput.\ Phys.\ Commun.\  {\bf 176} (2007) 383]
  
\bibitem{Rindlisbacher:2015pea}
  T.~Rindlisbacher and P.~de Forcrand,
  arXiv:1509.00087 [hep-lat].
  
\end{thebibliography}
\end{document}